\documentclass[conference]{IEEEtran}
\IEEEoverridecommandlockouts
\usepackage{cite}
\usepackage{amsmath,amssymb,amsfonts}
\usepackage{amsmath}
\DeclareMathOperator*{\argmax}{argmax}
\usepackage{graphicx}
\usepackage{textcomp}
\usepackage{xcolor}
\usepackage{graphicx}
\usepackage{float} 
\usepackage{subfigure}
\usepackage{amsmath}
\usepackage{amsfonts,amssymb}
\usepackage{mathrsfs}
\usepackage{mathtools}
\usepackage{algorithm}
\usepackage{algorithmicx}
\usepackage{algpseudocode}
\bibliography{IEEEabrv}
\def\BibTeX{{\rm B\kern-.05em{\sc i\kern-.025em b}\kern-.08em
    T\kern-.1667em\lower.7ex\hbox{E}\kern-.125emX}}

\begin{document}

\title{Secrecy Performance of Antenna-Selection-Aided MIMOME Channels with BPSK/QPSK Modulations \\
}

\author{\IEEEauthorblockN{Chongjun Ouyang, Zeliang Ou, Lu Zhang, Hongwen Yang and Xin Zhang}
\IEEEauthorblockA{\textit{Wireless Theories and Technologies Lab} \\
\textit{Beijing University of Posts and Telecommunications}\\
Beijing, China \\
\{DragonAim, ouzeliang, zhangl\_96, yanghong, zhangxin\}@bupt.edu.cn}
}

\maketitle

\begin{abstract}
This paper studies the secrecy performance of multiple-input multiple-output (MIMO) wiretap channels, also termed as multiple-input multiple-output multiple-eavesdropper (MIMOME) channels, under transmit antenna selection (TAS) and BPSK/QPSK modulations. In the main channel between the transmitter and the legitimate receiver, a single transmit antenna is selected to maximizes the instantaneous Signal to
Noise Ratio (SNR) at the receiver. At the receiver and the eavesdropper, selection combination (SC) is utilized. By assuming Rayleigh flat fading, we first derive the closed-form approximated expression for the ergodic secrecy rate when the channel state information of the eavesdropper (CSIE) is available at the transmitter. Next, analytical formulas for the approximated and asymptotic secrecy outage probability (SOP) are also developed when CSIE is unavailable. Besides theoretical derivations, simulation results are provided to demonstrate the approximation precision of the derived results. Furthermore, the asymptotic results reveal that the secrecy diversity order degrades into 0 due to the finite-alphabet inputs, which is totally different from that driven by the Gaussian inputs.

\end{abstract}

\begin{IEEEkeywords}
MIMOME channel, physical layer security, transmit antenna selection, BPSK/QPSK
\end{IEEEkeywords}

\section{Introduction}
\label{section1}
Physical layer (PHY) security has become a pivotal and pervasive concern in wireless communications due to its remarkable performance in information security enhancement. Different from the traditional cryptographic techniques \cite{b2}, physical layer security utilizes the inherent characteristics of wireless channels to ensure reliable transmission. In recent years, there has been an increasing interest in multiple-input multiple-output (MIMO) wiretap channels, also referred to as multiple-input
multiple-output multiple-eavesdropper (MIMOME) channels, where multiple antennas are deployed at the transmitter, the legitimate receiver and the eavesdropper. 

The secrecy capacity of MIMOME channels has been investigated in \cite{b4,b5} from an information-theoretic perspective. Later, theses works were further extended to large-scale systems in \cite{b7} to
declare the significant improvements of transmission security and reliability in massive MIMOME channels compared to the small-scale one. Nevertheless, the deployment of multiple antennas will result in high hardware cost since each antenna should be connected with an expensive Radio-Frequency (RF) chain. To settle this challenge, antenna selection (AS) \cite{b8} can be applied into the MIMO wiretap channels, which can alleviate the requirement on RF chains by selecting a subset of antennas to transceive signals. 

In the past years, many researches on antenna-selection-aided MIMOME channels were presented \cite{b12,b13,b14,b15,b18}. For example, Yang {\emph{et al.}}, in \cite{b12} and \cite{b13} derived closed-form expressions for the secrecy outage probability (SOP) in MIMOME channels under transmit antenna selection (TAS) with or without the impact of antenna correlation, respectively. Zhu {\emph{et al.}}, in \cite{b14} studied the probability of zero secrecy capacity for two TAS schemes depending on the availability of the eavesdropper's channel state information (CSI). Later, the achievable secrecy performance of MIMO wiretap channels in the presence of imperfect CSI was analyzed in \cite{b15}. In contrast to the explicit analysis in \cite{b12,b13,b14,b15}, Asaad {\emph{et al.}}, in \cite{b18} proposed asymptotically approximated results for both the average secrecy rate and the SOP in the limit of large-scale MIMOME systems under the norm-based TAS protocol. However, all these aforementioned works focused on the Gaussian input assumption. In fact, an important scenario which is necessary to be investigated when moving towards a practical implementation is the case where the channel inputs are constrained by finite constellation size.

Motivated by this, the works in \cite{b33} and \cite{b21} firstly studied the impacts of standard constellations on the achievable secrecy rates of Gaussian wiretap channels. Recently, most literatures about the MIMOME channels driven by finite-alphabet inputs focused on optimal precoding schemes \cite{b22,b24}, artificial noise (AN) design \cite{b28,b29} or the joint precoding and AN design \cite{b30}, while previous studies have not yet treated the TAS-aided MIMOME channels with finite-alphabet inputs in detail. Consequently, there is an urgent need to address the secrecy performance in TAS-aided MIMOME channels with inputs drawn form discrete constellations.

This paper detailedly analyzed the secrecy performance for MIMOME channels with transmit antenna selection and finite-alphabet inputs. To the best of our knowledge, this is the first time to propose a comprehensive theoretical analysis for the TAS-aided MIMOME channels with finite-alphabet inputs. For simplicity, assume that the modulation mode is BPSK/QPSK and the channel side information of the legitimate receiver (CSIL) is available at the transmitter. 
Closed-form approximated expressions for the ergodic secrecy rate and secrecy outage probability are respectively formulated in two scenarios: 1) For Scenario A: the eavesdropper's channel side information is unavailable at the transmitter (NCSIE), and 2) For Scenario B: the eavesdropper's channel side information (CSIE) is available. In each scenario, the derivations meet accurately the results given via numerical simulations. Finally, we set the Signal to Noise Ratio (SNR) of the main channel to infinity and derive the asymptotic secrecy outage probability. On the basis of the asymptotic SOP, we demonstrate that the secrecy diversity order \cite{b12} for the finite-alphabet inputs is totally different with that driven by Gaussian inputs due to the constraint imposed by the discrete signaling inputs. 

The remaining parts of this manuscript is structured as follows: Section \ref{section2} describes the system model. In Section \ref{section3}, the ergodic secrecy rate and secrecy outage probability of the MIMOME channel are investigated. The simulation results and corresponding analysis are shown in Section \ref{section4}. Finally, Section \ref{section5} concludes the paper.    


\section{System Model}
\label{section2}
In this paper, we consider a MIMO wiretap channel, where the transmitter, the legitimate receiver and the eavesdropper are equipped with $N_{\rm{A}}$, $N_{\rm{B}}$ and $N_{\rm{E}}$ antennas, respectively. Let ${\bf{H}}_{\rm{B}}$ and ${\bf{H}}_{\rm{E}}$ denote the main channel and the eavesdropper's channel, both of which are suffering independent and identical distributed (i.i.d.) Rayleigh flat fading with Gaussian noise. For the sake of brevity, suppose that the elements in the channel matrix ${\bf{H}}_{\rm{B}}\in{\mathbb{C}}^{N_{\rm{B}}\times N_{\rm{A}}}$ and  ${\bf{H}}_{\rm{E}}\in{\mathbb{C}}^{N_{\rm{E}}\times N_{\rm{A}}}$ are i.i.d. complex Gaussian random variables following ${\mathcal{CN}}\left(0,1\right)$.

Then, consider single antenna selection at the transmitter and selection combination (SC) at the legitimate receiver, which is a typical scenario declared in \cite{b12}. As a result, a single transmit/receive antenna pair between the transmitter and the legitimate receiver is selected to maximize the instantaneous SNR at the legitimate receiver. Consequently, the index of the selected antenna at the transmitter is given by
\begin{equation}
\beta^{\ast}=\argmax_{1\leq\alpha\leq N_{\rm{B}},1\leq\beta\leq N_{\rm{A}}}{\left|h_{{\rm{B}},\alpha,\beta}\right|},
\end{equation}
where $h_{{\rm{B}},\alpha,\beta}$ represents the element in ${\bf{H}}_{\rm{B}}$ located in the $\alpha$-th row and $\beta$-th line. Therefore, the magnitude of the main channel is $\left|f_{\rm{AB}}\right|=\max_{1\leq\alpha\leq N_{\rm{B}}}{\left|h_{{\rm{B}},\alpha,\beta^{\ast}}\right|}$. Furthermore, we assume that the eavesdropper also performs the SC to receive the secret message, thus the magnitude of the eavesdropper's channel is given by $\left|f_{\rm{AE}}\right|=\max_{1\leq\xi\leq N_{\rm{E}}}{\left|h_{{\rm{E}},\xi,\beta^{\ast}}\right|}$, where $h_{{\rm{E}},\xi,\beta}$ represents the element in ${\bf{H}}_{\rm{E}}$.

\subsection{Main channel}
After the antenna selection and SC, the received signal at the legitimate receiver can be written as
\begin{equation}
y_{\rm{B}}=\sqrt{\bar\gamma_{\rm{b}}}f_{\rm{AB}}x+n_{\rm{B}},
\end{equation}
where $x$ is the transmitted symbol constrained by finite alphabet with unit power, such as BPSK, $\bar\gamma_{\rm{b}}$ is the average per-antenna SNR of the main channel and $n_{\rm{B}}\in{\mathcal{CN}}\left(0,1\right)$ is the additive white Gaussian noise (AWGN). Let $\gamma_{\rm{b}}$ denote the instantaneous SNR at the legitimate receiver, then its probability density function (PDF) is given by
\begin{equation}
\label{eq3}
f_{\rm{b}}\left(\gamma_{\rm{b}}\right)=N_{\rm{A}}N_{\rm{B}}\left(1-{\rm{e}}^{-\frac{\gamma_{\rm{b}}}{\bar\gamma_{\rm{b}}}}\right)^{N_{\rm{A}}N_{\rm{B}}-1}{\rm{e}}^{-\frac{\gamma_{\rm{b}}}{\bar\gamma_{\rm{b}}}}\frac{1}{\bar\gamma_{\rm{b}}}.
\end{equation}

\subsection{Eavesdropper's channel}
The received signal at the eavesdropper is written as
\begin{equation}
y_{\rm{E}}=\sqrt{\bar\gamma_{\rm{e}}}f_{\rm{AE}}x+n_{\rm{E}},
\end{equation}
where $\bar\gamma_{\rm{e}}$ is the average per-antenna SNR of the eavesdropper's channel and $n_{\rm{E}}\in{\mathcal{CN}}\left(0,1\right)$ is the additive complex Gaussian noise. Since the eavesdropper only utilizes the antenna corresponding to the largest SNR, the PDF of the instantaneous SNR $\gamma_{\rm{e}}$ is 
\begin{equation}
\label{eq5}
f_{\rm{e}}\left(\gamma_{\rm{e}}\right)=N_{\rm{E}}\left(1-{\rm{e}}^{-\frac{\gamma_{\rm{e}}}{\bar\gamma_{\rm{e}}}}\right)^{N_{\rm{E}}-1}{\rm{e}}^{-\frac{\gamma_{\rm{e}}}{\bar\gamma_{\rm{e}}}}\frac{1}{\bar\gamma_{\rm{e}}}.
\end{equation}

\subsection{Achievable Secrecy Rate}
Since QPSK is the superposition of two orthogonal BPSK modulations, it is sufficient to consider BPSK.
Moreover, Shannon formula $\log_{2}\left(1+{\rm{SNR}}\right)$ can not be used for the input signals do not follow the Gaussian distribution. Assume that the transmitted data stream are i.i.d.
zero-mean binary symbols with equal probabilities, the input-output mutual information (MI) in terms of the SNR $\gamma$ under BPSK modulation over AWGN channels is formulated as \cite{b25}
\begin{equation}
\label{eq6}
{\mathcal{I}}\left(\gamma\right)=1-\int_{-\infty}^{+\infty}{\frac{1}{\sqrt{2\pi}}{\rm{e}}^{-\frac{u^2}{2}}\log_2\left(1+{\rm{e}}^{-2\sqrt{\gamma}u-2\gamma}\right){\rm{d}}u},
\end{equation} 
Therefore, the achievable secrecy rate of the wiretap channel with BPSK modulation is given by \cite{b27}
\begin{equation}
\label{eq7}
C_{\rm{s}}={\mathcal{I}}_{\rm{s}}\left(\gamma_{\rm{b}},\gamma_{\rm{e}}\right)=
\begin{cases}
{\mathcal{I}}\left(\gamma_{\rm{b}}\right)-{\mathcal{I}}\left(\gamma_{\rm{e}}\right),& {\gamma_{\rm{b}}>\gamma_{\rm{e}}}\\
0,& {\gamma_{\rm{b}}\leq\gamma_{\rm{e}}}
\end{cases}
\end{equation} 

\section{Secrecy Performance Analysis}
\label{section3}
This section will investigate the secrecy performance of the MIMOME channel under TAS in detail. In the followings, consider two scenarios stated before depending on whether the CSIE is available or not, and propose corresponding performance measurement metric for these cases respectively.
\subsection{CSIE}
\label{section3a}
\begin{figure}[!t] 
\setlength{\abovecaptionskip}{-4pt} 
\centering 
\includegraphics[width=0.45\textwidth]{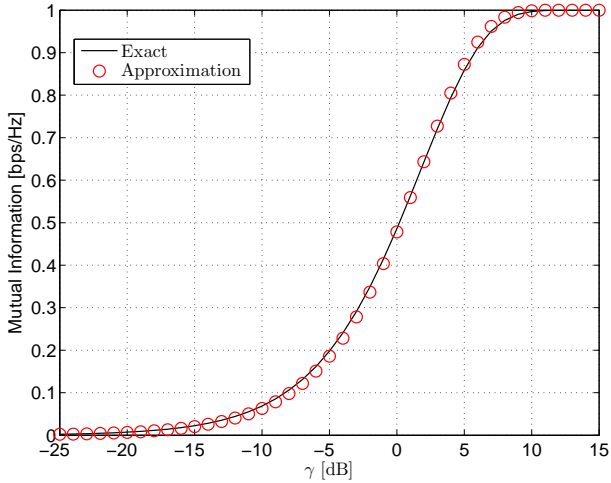} 
\caption{Comparison of exact and approximated MI of BPSK. The exact and approximated results are calculated by Equ. \eqref{eq6} and Equ. \eqref{eq9}, respectively.}
\label{figure}
\vspace{-4pt}
\end{figure}

When the CSIE is available, the ergodic secrecy rate, is usually utilized to measure the secrecy performance of the wiretap channel \cite{b27}, which is formulated as follows:
\begin{equation}
\label{eq8}
\begin{split}
{\bar{C}}_{\rm{s}}
=&{\mathbb{E}}\left[{\mathcal{I}}_{\rm{s}}\left(\gamma_{\rm{b}},\gamma_{\rm{e}}\right)\right]\\
=&\int_{0}^{+\infty}\int_{0}^{+\infty}{\mathcal{I}}_{\rm{s}}\left(\gamma_{\rm{b}},\gamma_{\rm{e}}\right)f_{\rm{b}}\left(\gamma_{\rm{b}}\right)f_{\rm{e}}\left(\gamma_{\rm{e}}\right){\rm{d}}\gamma_{\rm{b}}{\rm{d}}\gamma_{\rm{e}}\\
=&\int_{0}^{+\infty}\int_{\gamma_{\rm{e}}}^{+\infty}\left({\mathcal{I}}\left(\gamma_{\rm{b}}\right)-{\mathcal{I}}\left(\gamma_{\rm{e}}\right)\right)f_{\rm{b}}\left(\gamma_{\rm{b}}\right)f_{\rm{e}}\left(\gamma_{\rm{e}}\right){\rm{d}}\gamma_{\rm{b}}{\rm{d}}\gamma_{\rm{e}}.
\end{split}
\end{equation}
Nevertheless, due to the complexity of the expression for the MI in Equ. \eqref{eq6}, the exact value for the ergodic secrecy rate is hard to calculate. Fortunately, there exists a closed-form approximated formula for the MI, with a compact form, which is written as:
\begin{equation}
\label{eq9}
{\mathcal{I}}\left(\gamma\right)\approx1-{\rm{e}}^{-\phi\gamma},
\end{equation}  
where $\phi=0.6507$. To examine the approximation precision of Equ. \eqref{eq9}, {\figurename} \ref{figure} compares the exact and approximated input-ouput MI of BPSK in terms of the SNR. As can be seen from this figure, the approximation effect is fantastic for all SNR ranges. Additionally, as $\gamma$ tending to be 0 or $+\infty$, $\left(1-{\rm{e}}^{-\phi\gamma}\right)$ will tend to be 0 or 1, which can accord with practice. In summary, these facts indicate that it is accurate enough to estimate the exact secrecy rate using this formula, which is given by
\begin{equation}
\label{EQ10}
C_{\rm{s}}={\mathcal{I}}_{\rm{s}}\left(\gamma_{\rm{b}},\gamma_{\rm{e}}\right)\approx
\begin{cases}
{\rm{e}}^{-\phi\gamma_{\rm{e}}}-{\rm{e}}^{-\phi\gamma_{\rm{b}}},& {\gamma_{\rm{b}}>\gamma_{\rm{e}}}\\
0,& {\gamma_{\rm{b}}\leq\gamma_{\rm{e}}}.
\end{cases}
\end{equation}

Substituting Equ. \eqref{EQ10} into Equ. \eqref{eq8}, the expression for the approximated ergodic secrecy rate can be developed, that is
\begin{equation}
\label{eq10}
\begin{split}
{\bar{C}}_{\rm{s}}\approx&\int_{0}^{+\infty}\int_{\gamma_{\rm{e}}}^{+\infty}
\left({\rm{e}}^{-\phi\gamma_{\rm{e}}}-{\rm{e}}^{-\phi\gamma_{\rm{b}}}\right)f_{\rm{b}}\left(\gamma_{\rm{b}}\right)f_{\rm{e}}\left(\gamma_{\rm{e}}\right){\rm{d}}\gamma_{\rm{b}}{\rm{d}}\gamma_{\rm{e}}\\
=&\underbrace{\int_{0}^{+\infty}\int_{\gamma_{\rm{e}}}^{+\infty}
{\rm{e}}^{-\phi\gamma_{\rm{e}}}f_{\rm{b}}\left(\gamma_{\rm{b}}\right)f_{\rm{e}}\left(\gamma_{\rm{e}}\right){\rm{d}}\gamma_{\rm{b}}{\rm{d}}\gamma_{\rm{e}}}_{\Psi_1}\\
&-\underbrace{\int_{0}^{+\infty}\int_{\gamma_{\rm{e}}}^{+\infty}
{\rm{e}}^{-\phi\gamma_{\rm{b}}}f_{\rm{b}}\left(\gamma_{\rm{b}}\right)f_{\rm{e}}\left(\gamma_{\rm{e}}\right){\rm{d}}\gamma_{\rm{b}}{\rm{d}}\gamma_{\rm{e}}}_{\Psi_2}.
\end{split}
\end{equation}
Using Equ. \eqref{eq3} and Equ. \eqref{eq5}, the approximated ergodic secrecy rate can be derived as
\begin{equation}
\label{eq3_1}
\begin{split}
{\bar{C}}_{\rm{s}}
\approx
&\sum_{k=0}^{N_{\rm{A}}N_{\rm{B}}-1}\sum_{j=0}^{N_{\rm{E}}-1}\frac{N_{\rm{A}}N_{\rm{B}}N_{\rm{E}}}{\bar\gamma_{\rm{e}}}\binom{N_{\rm{A}}N_{\rm{B}}-1}{k}\binom{N_{\rm{E}}-1}{j}\\
&\times\frac{\left(-1\right)^{k+j}}{\phi+\frac{j+1}{\bar\gamma_{\rm{e}}}+\frac{k+1}{\bar\gamma_{\rm{b}}}}\left(\frac{1}{k+1}-\frac{1}{k+1+\phi{\bar\gamma_{\rm{b}}}}\right).
\end{split}
\end{equation}
We first look into $\Psi_1$ and apply \cite[Equ. (1.111)]{b26} for the binomial expansion into Equ. \eqref{eq3} and Equ. \eqref{eq5}, then $\Psi_1$ can be written as  
\begin{equation}
\begin{split}
\Psi_1=&\frac{N_{\rm{A}}N_{\rm{B}}N_{\rm{E}}}{{\bar\gamma_{\rm{b}}}{\bar\gamma_{\rm{e}}}}\int_{0}^{+\infty}\left(1-{\rm{e}}^{-\frac{\gamma_{\rm{e}}}{\bar\gamma_{\rm{e}}}}\right)^{N_{\rm{E}}-1}{\rm{e}}^{-\frac{\gamma_{\rm{e}}}{\bar\gamma_{\rm{e}}}}{\rm{e}}^{-\phi\gamma_{\rm{e}}}{\rm{d}}\gamma_{\rm{e}}\\
&\times\int_{\gamma_{\rm{e}}}^{+\infty}\left(1-{\rm{e}}^{-\frac{\gamma_{\rm{b}}}{\bar\gamma_{\rm{b}}}}\right)^{N_{\rm{A}}N_{\rm{B}}-1}{\rm{e}}^{-\frac{\gamma_{\rm{b}}}{\bar\gamma_{\rm{b}}}}{\rm{d}}\gamma_{\rm{b}}\\
=&\sum_{k=0}^{N_{\rm{A}}N_{\rm{B}}-1}\sum_{j=0}^{N_{\rm{E}}-1}\frac{N_{\rm{A}}N_{\rm{B}}N_{\rm{E}}}{\bar\gamma_{\rm{b}}\bar\gamma_{\rm{e}}}\binom{N_{\rm{A}}N_{\rm{B}}-1}{k}\binom{N_{\rm{E}}-1}{j}\\
&\times\underbrace{\int_{0}^{+\infty}{\rm{e}}^{-\left(\phi+\frac{j+1}{\bar\gamma_{\rm{e}}}\right)\gamma_{\rm{e}}}{\rm{d}}\gamma_{\rm{e}}\int_{\gamma_{\rm{e}}}^{+\infty}{\rm{e}}^{-\left(\frac{j+1}{\bar\gamma_{\rm{b}}}\right)\gamma_{\rm{b}}}{\rm{d}}\gamma_{\rm{b}}}_{\Psi_3\left(j,k\right)},
\end{split}
\end{equation}
in which ${\Psi_3}\left(j,k\right)$ is easy to solve for it only contains exponential functions. On the other hand, $\Psi_2$ can be also calculated following the similar steps as $\Psi_1$. Next, substituting $\Psi_1$ and $\Psi_2$ into Equ. \eqref{eq10} and performing some basic mathematical manipulations, the final result in Equ. \eqref{eq3_1} can be derived. 

This subsection has detailedly investigated the scenario when CSIE is available at the transmitter and the approximated closed-form expression for the ergodic secrecy rate is derived. The next subsection, therefore, moves on to discuss the situation when the transmitter knows nothing about the CSIE.

\subsection{NCSIE}
\label{section3b}
When the CSI of the eavesdropper is unknown at the transmitter, the secrecy rate defined in Equ. \eqref{eq7} is not achievable \cite{b27}. Under this circumstance, the secrecy performance is usually measured by the secrecy outage probability (SOP), which is defined as 
\begin{equation}
\label{eq12}
P_{\rm{out}}\left(R_{\rm{s}}\right)={\rm{Pr}}\left(C_{\rm{s}}<R_{\rm{s}}\right),
\end{equation}
where $R_{\rm{s}}\geq0$ represents the preset secrecy rate. According to the definition of SOP, it is clear that SOP denotes the probability that the achievable secrecy rate is less than a predetermined secrecy transmission rate, below which secure transmission is not guaranteed.
\subsubsection{Probability of non-zero secrecy}
Before explaining the general SOP (for $R_{\rm{s}}>0$), we first examine the condition of positive secrecy rate i.e., the probability of taking positive values for the secrecy rate $C_{\rm{s}}$. On the basis of Equ. \eqref{eq7} and Equ. \eqref{eq12}, the probability of non-zero secrecy rate is formulated as
\begin{equation}
\label{eq13}
\begin{split}
{\rm{Pr}}\left(C_{\rm{s}}>0\right)&=1-P_{\rm{out}}\left(0\right)={\rm{Pr}}\left(\gamma_{\rm{b}}>\gamma_{\rm{e}}\right)\\
&=\int_{0}^{+\infty}\int_{\gamma_{\rm{e}}}^{+\infty}f_{\rm{b}}\left(\gamma_{\rm{b}}\right)f_{\rm{e}}\left(\gamma_{\rm{e}}\right){\rm{d}}\gamma_{\rm{b}}{\rm{d}}\gamma_{\rm{e}}.
\end{split}
\end{equation}

By substituting Equ. \eqref{eq3} and Equ. \eqref{eq5} into Equ. \eqref{eq13} and calculating the resultant integrals on the basis of \cite[Equ. (3.432.1)]{b26}, the closed-form expression for the probability of positive secrecy rate can be obtained, namely
\begin{equation}
\label{eq3_2}
\begin{split}
{\rm{Pr}}\left(C_{\rm{s}}>0\right)=&\sum_{k=0}^{N_{\rm{A}}N_{\rm{B}}-1}\sum_{j=0}^{N_{\rm{E}}-1}\frac{\left(-1\right)^{k+j}}{\left(k+1\right)\bar\gamma_{\rm{e}}}\frac{N_{\rm{A}}N_{\rm{B}}N_{\rm{E}}}{\frac{j+1}{\bar\gamma_{\rm{e}}}+\frac{k+1}{\bar\gamma_{\rm{b}}}}\\
&\times\binom{N_{\rm{A}}N_{\rm{B}}-1}{k}\binom{N_{\rm{E}}-1}{j}.
\end{split}
\vspace*{-4pt}
\end{equation}

\subsubsection{General secrecy outage probability}On the basis of the definition of the secrecy rate and the secrecy outage probability, the expression of the SOP is given by
\begin{equation}
\begin{split}
P_{\rm{out}}\left(R_{\rm{s}}\right)
=&{\rm{Pr}}\left(C_{\rm{s}}<R_{\rm{s}}|\gamma_{\rm{b}}>\gamma_{\rm{e}}\right){\rm{Pr}}\left(\gamma_{\rm{b}}>\gamma_{\rm{e}}\right)\\
&+{\rm{Pr}}\left(C_{\rm{s}}<R_{\rm{s}}|\gamma_{\rm{b}}<\gamma_{\rm{e}}\right){\rm{Pr}}\left(\gamma_{\rm{b}}<\gamma_{\rm{e}}\right).
\end{split}
\end{equation}
By Equ. \eqref{eq7}, $C_{\rm{s}}$ is 0 when $\gamma_{\rm{b}}<\gamma_{\rm{e}}$; on the other hand, the preset $R_{\rm{s}}$ is positive, thus ${\rm{Pr}}\left(C_{\rm{s}}<R_{\rm{s}}|\gamma_{\rm{b}}<\gamma_{\rm{e}}\right)$ equals to 1. As a result, the SOP can be simplified as
\begin{equation}
\label{EQ18}
\begin{split}
P_{\rm{out}}\left(R_{\rm{s}}\right)
=&{\rm{Pr}}\left(C_{\rm{s}}<R_{\rm{s}}|\gamma_{\rm{b}}>\gamma_{\rm{e}}\right)\\
&\times{\rm{Pr}}\left(\gamma_{\rm{b}}>\gamma_{\rm{e}}\right)+{\rm{Pr}}\left(\gamma_{\rm{b}}<\gamma_{\rm{e}}\right).
\end{split}
\end{equation}
In Equ. \eqref{EQ18}, the final results for both ${\rm{Pr}}\left(\gamma_{\rm{b}}<\gamma_{\rm{e}}\right)$ and ${\rm{Pr}}\left(\gamma_{\rm{b}}>\gamma_{\rm{e}}\right)$ can be directly obtained using Equ. \eqref{eq3_2}. Next, let us turn to the term ${\rm{Pr}}\left(C_{\rm{s}}<R_{\rm{s}}|\gamma_{\rm{b}}>\gamma_{\rm{e}}\right)$. On the basis of \cite{b27}, ${\rm{Pr}}\left(C_{\rm{s}}<R_{\rm{s}}|\gamma_{\rm{b}}>\gamma_{\rm{e}}\right)$ can be expressed as  
\begin{equation}
\label{eq17}
\begin{split}
&{\rm{Pr}}\left(C_{\rm{s}}<R_{\rm{s}}|\gamma_{\rm{b}}>\gamma_{\rm{e}}\right)\\
=&\frac{1}{\Theta}\int_{0}^{+\infty}\int_{\gamma_{\rm{e}}}^{{\mathcal{I}}^{-1}\left(R_{\rm{s}}+{\mathcal{I}}\left(\gamma_{\rm{e}}\right)\right)}f_{\rm{b}}\left(\gamma_{\rm{b}}\right)f_{\rm{e}}\left(\gamma_{\rm{e}}\right){\rm{d}}\gamma_{\rm{b}}{\rm{d}}\gamma_{\rm{e}}\\
=&\frac{1}{\Theta}\underbrace{\int_{0}^{+\infty}\int_{0}^{{\mathcal{I}}^{-1}\left(R_{\rm{s}}+{\mathcal{I}}\left(\gamma_{\rm{e}}\right)\right)}f_{\rm{b}}\left(\gamma_{\rm{b}}\right)f_{\rm{e}}\left(\gamma_{\rm{e}}\right){\rm{d}}\gamma_{\rm{b}}{\rm{d}}\gamma_{\rm{e}}}_{{\Psi}_4}\\
&-\frac{1}{\Theta}\underbrace{\int_{0}^{+\infty}\int_{0}^{\gamma_{\rm{e}}}f_{\rm{b}}\left(\gamma_{\rm{b}}\right)f_{\rm{e}}\left(\gamma_{\rm{e}}\right){\rm{d}}\gamma_{\rm{b}}{\rm{d}}\gamma_{\rm{e}}}_{{{\Psi}}_5},
\end{split}
\end{equation}
where $\Theta={\rm{Pr}}\left(\gamma_{\rm{b}}>\gamma_{\rm{e}}\right)$ and ${\mathcal{I}}^{-1}\left(\cdot\right)$ denotes the inverse function of ${\mathcal{I}}\left(\cdot\right)$. Note that $\Psi_{5}=1-{\rm{Pr}}\left(\gamma_{\rm{b}}>\gamma_{\rm{e}}\right)$ which can be simply solved by Equ. \eqref{eq3_2}. As for $\Psi_4$, the approximated formula $\left(1-\rm{e}^{-\phi\gamma}\right)$ can be utilized to simplify its calculation, since the inverse function of ${\mathcal{I}}\left(\gamma\right)$ is difficult to derive. Consequently,
\begin{equation}
\label{eq18} 
\Psi_{4}\approx\int_{0}^{+\infty}\int_{0}^{-\ln\left({\rm{e}}^{-\phi\gamma_{\rm{e}}}-R_{\rm{s}}\right)/\phi}f_{\rm{b}}\left(\gamma_{\rm{b}}\right)f_{\rm{e}}\left(\gamma_{\rm{e}}\right){\rm{d}}\gamma_{\rm{b}}{\rm{d}}\gamma_{\rm{e}}
\end{equation}
We substitute Equ. \eqref{eq3} and Equ. \eqref{eq5} into Equ. \eqref{eq18} and expand the PDF by the binomial expansion. After some lines of derivation, the approximated result for $\Psi_{4}$ is given by
\begin{equation}
\label{eq19} 
\begin{split}
&\Psi_{4}\approx\\
&1-\sum_{k=0}^{N_{\rm{A}}N_{\rm{B}}-1}\sum_{j=0}^{N_{\rm{E}}-1}\frac{\left(-1\right)^{k+j}}{k+1}\frac{N_{\rm{A}}N_{\rm{B}}N_{\rm{E}}}{{\bar{\gamma}_{\rm{e}}}\phi}\binom{N_{\rm{A}}N_{\rm{B}}-1}{k}\\
&\times\binom{N_{\rm{E}}-1}{j}\left(-R_{\rm{s}}\right)^{v_k}{\rm{B}}\left(u_j,1\right){_2F_1}\left(-v_k,u_j;u_j+1;\frac{1}{R_{\rm{s}}}\right),
\end{split}
\end{equation}
in which $v_k=\frac{1+k}{\bar{\gamma}_{\rm{b}}\phi}$ and $u_j=\frac{1+j}{\bar{\gamma}_{\rm{e}}\phi}$. Besides, ${\rm{B}}\left(\cdot,\cdot\right)$ and $_2F_1\left(\cdot,\cdot;\cdot;\cdot\right)$ denote the Beta function \cite[Equ. (8.380)]{b26} and Gauss hypergeometric function \cite[Equ. (9.100)]{b26}, respectively.
Afterwards, substituting the results of $\Psi_4$ and  $\Psi_5$ into Equ. \eqref{eq17} and performing some mathematical manipulations, the approximated expression of the SOP can be obtained, which is exhibited on the top of next page.
 
\begin{figure*}
\setlength{\abovecaptionskip}{-4pt} 
\begin{equation}
\label{eq3_6}
\begin{split}
&{\rm{Pr}}\left(C_{\rm{s}}<R_{\rm{s}}\right)
\approx
1-\sum_{k=0}^{N_{\rm{A}}N_{\rm{B}}-1}\sum_{j=0}^{N_{\rm{E}}-1}\frac{\left(-1\right)^{k+j+v_k}}{k+1}\frac{N_{\rm{A}}N_{\rm{B}}N_{\rm{E}}}{{\bar{\gamma}_{\rm{e}}}\phi}\binom{N_{\rm{A}}N_{\rm{B}}-1}{k}\binom{N_{\rm{E}}-1}{j}R_{\rm{s}}^{v_k}{\rm{B}}\left(u_j,1\right){_2F_1}\left(-v_k,u_j;u_j+1;\frac{1}{R_{\rm{s}}}\right)
\end{split}
\end{equation}
\hrulefill
\vspace*{-10pt}
\end{figure*}

Let $R_{\rm{s}}$ and $\bar\gamma_{\rm{e}}$ be fixed, and the secrecy outage probability will tend to be 0 as $\bar\gamma_{\rm{b}}\rightarrow+\infty$ when the input signals follow Gaussian distribution \cite{b12}, for the channel capacity of the main channel can increase monotonically without any limitation. In contrast with the Gaussian inputs, the maximal input-output MI of the main channel is a constant for the digital-modulation systems due to the constraint of finite constellation size. Next, we still take BPSK as an example and evaluate the asymptotic SOP when $\bar\gamma_{\rm{b}}\rightarrow+\infty$. By Equ. \eqref{eq6}, the MI for the main channel will tend to be 1 as $\bar\gamma_{\rm{b}}$ rises up, thus the asymptotic SOP can be written as
\begin{equation}
\label{eq3_7}
\begin{split}
{\rm{P}}^{\infty}_{\rm{out}}\left(R_{\rm{s}}\right)&={\rm{Pr}}\left(1-{\mathcal{I}}\left(\gamma_{\rm{e}}\right)<R_{\rm{s}}\right)
={\rm{Pr}}\left(\gamma_{\rm{e}}>{\mathcal{I}}^{-1}\left(1-R_{\rm{s}}\right)\right)\\
&=1-\left(1-{\rm{e}}^{-\frac{{\mathcal{I}}^{-1}\left(1-R_{\rm{s}}\right)}{\bar\gamma_{\rm{e}}}}\right)^{N_{\rm{E}}}.
\end{split}
\end{equation}  
Since ${\mathcal{I}}^{-1}\left(\cdot\right)$ is hard to solve, the approximated expression $\left(1-{\rm{e}}^{-\phi\gamma}\right)$ can be used to approximate the final result, thus the asymptotic SOP can be estimated as
\begin{equation}
\label{eq3_8}
{\rm{P}}^{\infty}_{\rm{out}}\left(R_{\rm{s}}\right)\approx1-\left(1-{R_{\rm{s}}}^{\frac{1}{0.6507\bar\gamma_{\rm{e}}}}\right)^{N_{\rm{E}}}.
\end{equation}
As can be seen from Equ. \eqref{eq3_8}, $\bar\gamma_{\rm{b}}$, $N_{\rm{A}}$ and $N_{\rm{B}}$ have no impact on the the asymptotic SOP. Nevertheless, on the basis of \cite{b12}, the asymptotic SOP for Gaussian inputs when $\bar\gamma_{\rm{b}}\rightarrow+\infty$ is 
\begin{equation}
{\rm{P}}^{\infty}_{\rm{out}}\left(R_{\rm{s}}\right)=\left(A\bar\gamma_{\rm{b}}\right)^{-G}+o\left(\bar\gamma_{\rm{b}}^{-G}\right),
\end{equation} 
where $A$ is only related with $\bar\gamma_{\rm{e}}$ and $N_{\rm{E}}$, $o\left(\cdot\right)$ denotes higher order terms and $G=N_{\rm{A}}N_{\rm{B}}$. According to \cite{b12}, $G$ is termed as secrecy diversity order, which represents the slope of the SOP curve. On the basis of our asymptotic result, the secrecy diversity order for finite-alphabet inputs has degraded from $N_{\rm{A}}N_{\rm{B}}$ into 0 for ${\rm{P}}^{\infty}_{\rm{out}}\left(R_{\rm{s}}\right)$ is irrelevant to $\bar\gamma_{\rm{b}}$, which is totally different from the scenario of Gaussian inputs. It is clear that the maximal MI for the main channel is a constant related with the modulation modes, which has nothing to do with $\bar\gamma_{b}$ due to the limitation of finite constellation size. And this is just the reason why ${\rm{P}}^{\infty}_{\rm{out}}\left(R_{\rm{s}}\right)$ is unaffected by $\bar\gamma_{b}$, causing the secrecy diversity order to be 0.

\section{Simulation Results}
\label{section4}
In this section, numerical and simulation results derived in the preceding sections are given. As stated before, there is no closed-form expression for the ergodic secrecy rate, thus Monte-Carlo simulations with a large number of trials are utilized to approach the exact value, which will be used to examine the feasibility and validity of the former derivations.

\begin{figure}[!t] 
\setlength{\abovecaptionskip}{-4pt} 
\centering 
\includegraphics[width=0.45\textwidth]{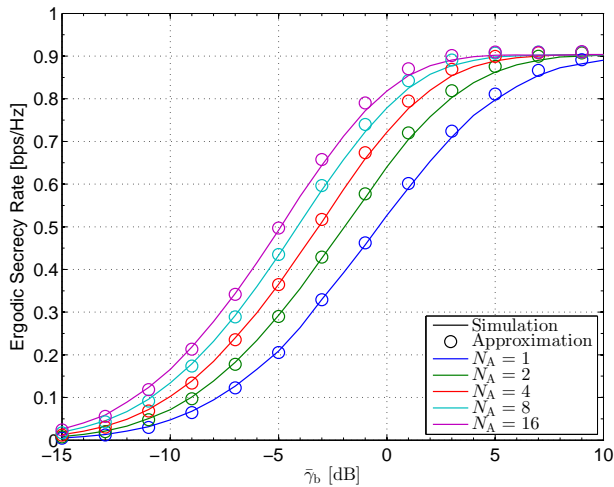} 
\caption{Simulated and approximated ($\circ$) ergodic secrecy capacity versus $\bar\gamma_{\rm{b}}$ for $N_{\rm{B}}=3$, $N_{\rm{E}}=2$ and $\bar\gamma_{\rm{e}}=-10$ dB.}
\label{figure3_2}
\vspace*{-4pt}
\end{figure}

\begin{figure}[!t] 
\setlength{\abovecaptionskip}{-4pt} 
\centering 
\includegraphics[width=0.45\textwidth]{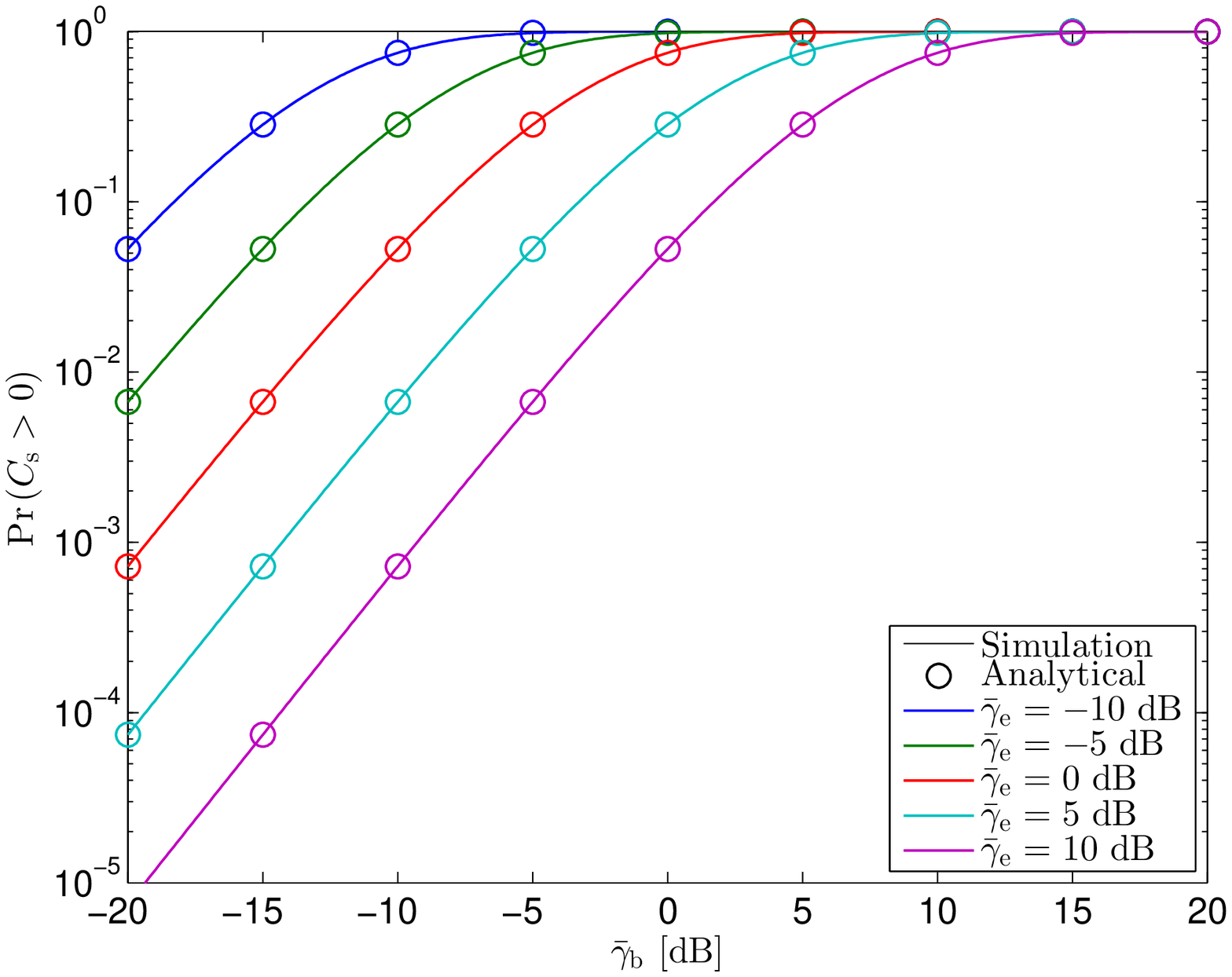} 
\caption{Simulated and analytical probability of non-zero secrecy capacity versus $\bar\gamma_{\rm{b}}$ for $N_{\rm{A}}=3$, $N_{\rm{B}}=N_{\rm{E}}=2$.}
\label{figure3_3}
\vspace*{-4pt}
\end{figure}

\begin{figure}[!t] 
\setlength{\abovecaptionskip}{-4pt} 
    \centering
    \subfigure[$R_{\rm{s}}=0.5$ bps/Hz, $\bar\gamma_{\rm{e}}=-10$ dB]
    {
        \includegraphics[width=0.45\textwidth]{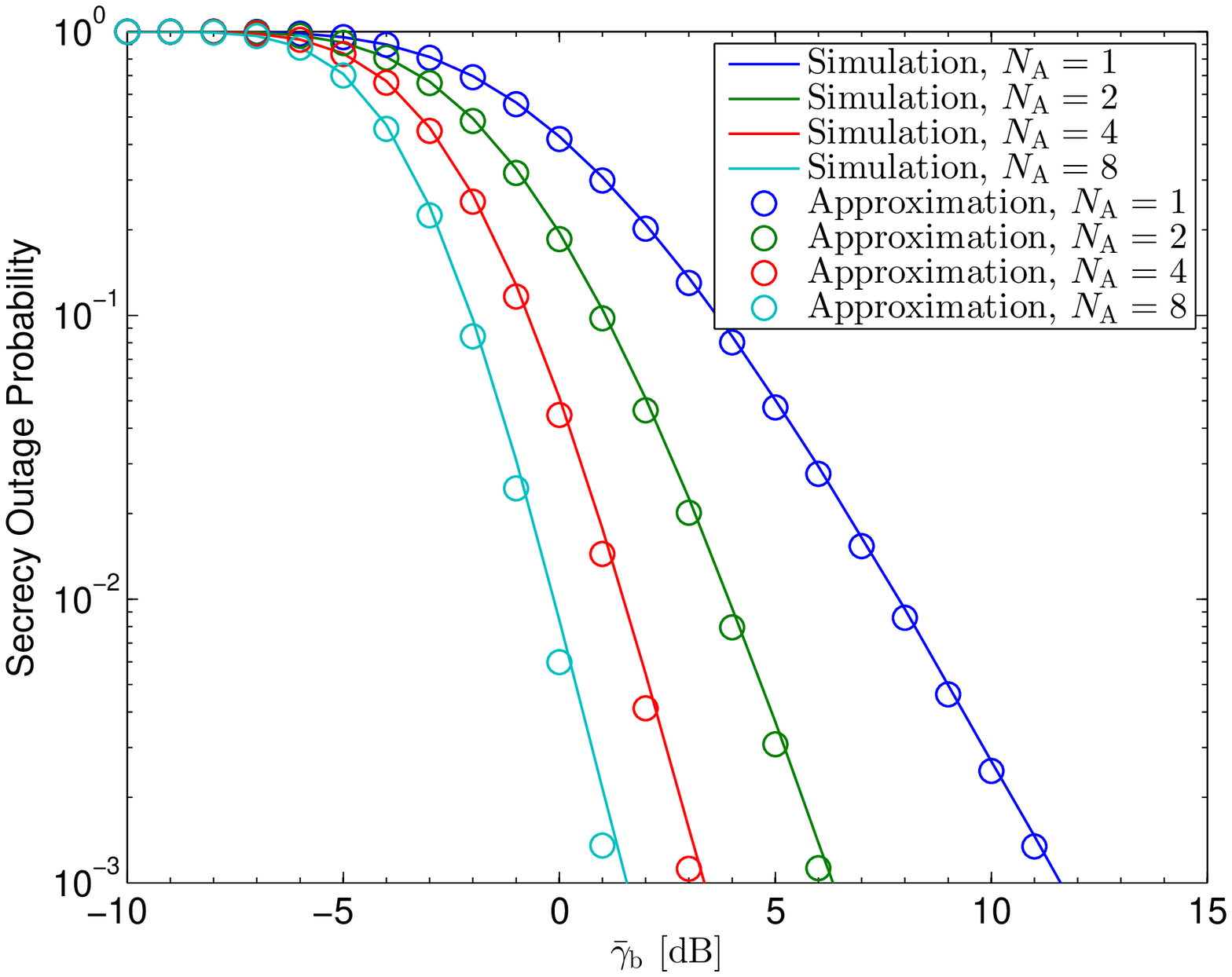}
	   \label{figure3_4a}	   
    } \\
   \subfigure[$R_{\rm{s}}=0.5$ bps/Hz, $\bar\gamma_{\rm{e}}=-6$ dB]
    {
        \includegraphics[width=0.45\textwidth]{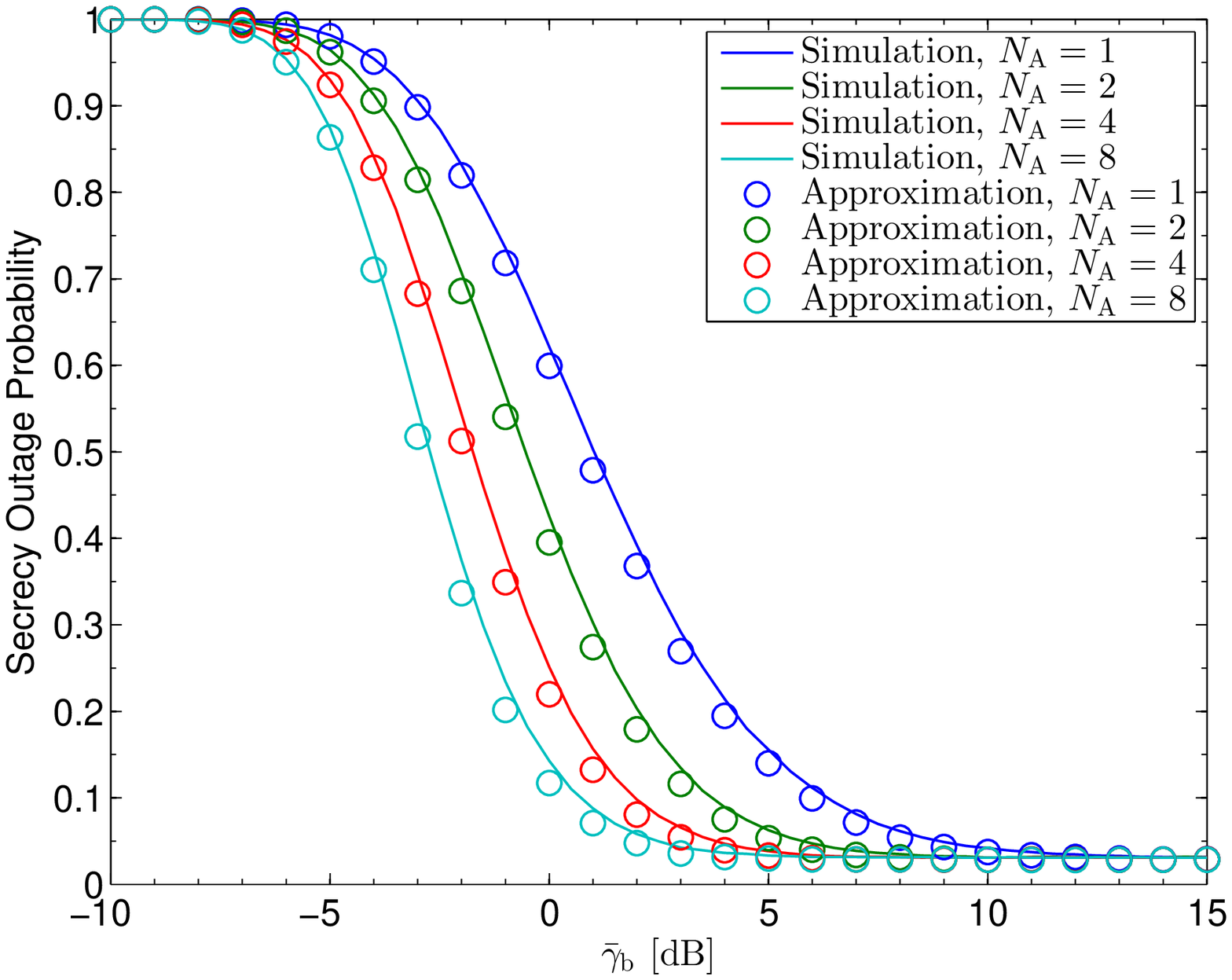}
	   \label{figure3_4b}	   
    } 
\\
    \caption{Simulated and approximated secrecy outage secrecy versus $\bar\gamma_{\rm{b}}$ for $N_{\rm{B}}=3$, $N_{\rm{E}}=2$.}
    \label{figure3_4}
	\vspace{-4pt}
\end{figure}

\begin{figure}[!t] 
\setlength{\abovecaptionskip}{-4pt} 
\centering 
\includegraphics[width=0.45\textwidth]{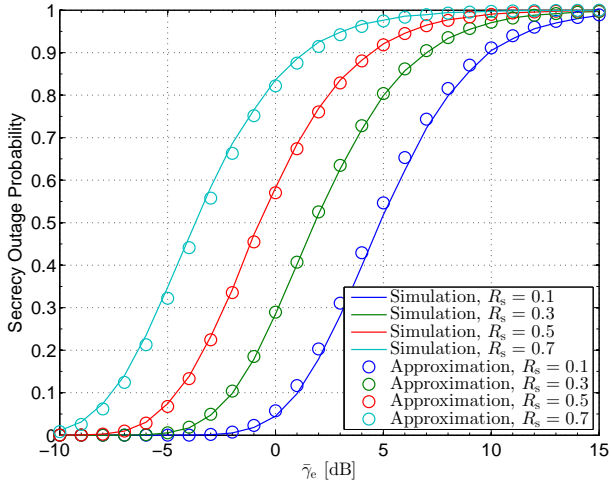} 
\caption{Simulated and approximated limit secrecy outage secrecy versus $\bar\gamma_{\rm{e}}$ for $\bar\gamma_{\rm{b}}=30$ dB, $N_{\rm{E}}=N_{\rm{B}}=2$ and $N_{\rm{A}}=5$.}
\label{figure3_5}
\vspace*{-4pt}
\end{figure}

Let us fix $N_{\rm{B}}=3$, $N_{\rm{E}}=2$ and increase $N_{\rm{A}}$ from 1 to 16. Besides, the average SNR at the eavesdropper is set to be $\bar\gamma_{\rm{e}}=-10$ dB as $\bar\gamma_{\rm{b}}$ ranges between -15 dB and 10 dB. {\figurename} \ref{figure3_2} presents the results of simulated and approximated ergodic secrecy rate, obtained by Monte-Carlo experiments and  Equ. \eqref{eq3_1} respectively. 
As shown in {\figurename} \ref{figure3_2}, the curves denoting the simulated results nearly coincide with the circles denoting the approximated results, which suggests that the former derivation about the ergodic secrecy rate in Section \ref{section3a} is correct. In addition, the coincidence of the curves also indicates that the approximated formula of the MI for BPSK serves as a efficient tool that can be utilized to simplify some derivations. Finally, comparing the curves for different $N_{\rm{A}}$, it can be observed that the larger the $N_{\rm{A}}$, the more secrecy transmission rate the antenna selection system can achieve even though the total number of the RF chains is fixed.     

Next, let us turn to the scenario when the CSI of the eavesdropper is unavailable at the transmitter. In the following part, the simulation and numerical results for the probability of non-zero secrecy rate will be presented first, then we will turn to the secrecy outage probability. {\figurename} \ref{figure3_3} compares the simulated and analytical probability of non-zero secrecy rate versus $\bar\gamma_{\rm{b}}$. The analytical results are calculated by Equ. \eqref{eq3_2} and the simulated results are obtained by Monte-Carlo experiments. To approach the exact probability of non-zero secrecy rate, the Monte-Carlo experiments consist of $10^{7}$ trails. It can be seen form {\figurename} \ref{figure3_3} that the analytical results meet accurately with the simulation results, which further verifies the former derivation. Furthermore, the probability decreases with the increment of $\bar\gamma_{\rm{e}}$ but increases as $\bar\gamma_{\rm{b}}$ rises up, which suggests the active and passive effect of the legitimate receiver and the eavesdropper in the wiretap channel. Note that the approximated formula of the mutual information is not utilized during the derivation of the probability of non-zero secrecy capacity, thus the derived results just represent the exact values of the probability. 

Then, {\figurename} \ref{figure3_4} provides the simulated and approximated results for the secrecy outage probability in terms of $\gamma_{\rm{b}}$ as $N_{\rm{A}}$ increases from 1 to 8 and $\gamma_{\rm{e}}$ ranges between -10 dB and -6 dB. As shown in {\figurename} \ref{figure3_4a} and {\figurename} \ref{figure3_4b}, the approximated SOP, calculated by Equ. \eqref{eq3_6}, agrees well with the simulated results, which supports our approximate derivations in Section \ref{section3b}. As explained in the previous section, the SOP can not tend to be 0 with the increment of $\bar\gamma_{\rm{b}}$ due to the limitation of finite-alphabet inputs, and this phenomenon can be clearly observed from {\figurename} \ref{figure3_4b}. Overall, taken the results in {\figurename} \ref{figure3_2} and {\figurename} \ref{figure3_4} together, it makes sense to apply the approximation expression $\left(1-{\rm{e}}^{-0.6507\gamma}\right)$ into the estimation of secrecy performance for wiretap channel with BPSK/QPSK modulations.

As stated before, the SOP can not decrease continuously with the increment of $\bar{\gamma}_{\rm{b}}$ when $\bar{\gamma}_{\rm{e}}$ is fixed. 
To further explore the asymptotic behavior of the wiretap channel, {\figurename} \ref{figure3_5} plots the asymptotic SOP versus $\bar\gamma_{\rm{e}}$. Moreover, $\bar{\gamma}_{\rm{b}}$ is fixed to 30 dB to make sure the mutual information of the main channel reaches 1. As can be seen from this graph, the asymptotic SOP will increase as $\bar{\gamma}_{\rm{e}}$ rises up. In addition, the approximated results and the simulated results nearly equal to each other, which verifies the validity of the deduction.

\section{Conclusion}
\label{section5}
This paper detailedly analyzes the secrecy performance of antenna-selection-aided MIMOME channels under BPSK/QPSK modulations. Approximated expressions for the ergodic secrecy rate and SOP are proposed and discussed in the situations when the CSIE is available or unavailable. Simulation shows that the approximated results indicated high precision and can serve as the estimation in the practical wiretap channel. Additionally, discussion about the SOP suggests that the finite alphabet input is the main limitation of the security and reliability in digital-modulation systems.

\vspace{12pt}
\end{document}